\documentclass{article}

\newenvironment{Presented}{\begin{quotation} \begin{center} \vspace{3in}
             PRESENTED AT\end{center}\bigskip 
      \begin{center}\begin{large}}{\end{large}\end{center} \end{quotation}}

\usepackage[dvips]{graphicx}
\usepackage{float}
\usepackage{url}
\usepackage{cite}
\usepackage{authblk}
\usepackage{amsmath,amssymb}

\title{\textbf{ATLAS Searches for Beyond the Standard Model Higgs Bosons}\\DPF2013-198}

\author{C.T. Potter}
\affil{Department of Physics, University of Oregon}

\date{\today}

\begin{document}

\maketitle
\thispagestyle{empty}

\begin{abstract}
The present status of ATLAS searches for Higgs bosons in extensions of the Standard Model (SM) is presented. This includes searches for the Higgs bosons of the Two-Higgs-Doublet Model (2HDM), the Minimal Supersymmetric Model (MSSM), the Next-to-Minimal Supersymmetric Model (NMSSM) and models with an invisibly decaying Higgs boson. A review of the phenomenology of the Higgs sectors of these models is given together with the search strategy and the resulting experimental constraints. 
\end{abstract}

\begin{Presented}
DPF 2013\\
The Meeting of the American Physical Society\\
Division of Particles and Fields\\
Santa Cruz, California, August 13--17, 2013\\
\end{Presented}

\newpage

\setcounter{page}{1}

\section{Introduction}

The discovery of a new 125 GeV boson by CMS \cite{Chatrchyan:2012ufa} and ATLAS \cite{Aad:2012tfa} at the Large Hadron Collider (LHC)  is a striking success for the Standard Model (SM) mechanism for electroweak symmetry breaking \cite{PhysRevLett.13.508, PhysRevLett.13.321, PhysRevLett.13.585}. But the SM cannot be a complete theory, as it does not explain gravity, dark matter, dark energy, or the baryon asymmetry. Moreover the SM suffers from the well known hierarchy problem: the SM Higgs boson mass is quadratically divergent.

This suggests that the SM must be embedded within a more complete theoretical model. Many such models beyond the SM (BSM) predict the existence of more than one Higgs boson. In a generic Two-Higgs-Doublet Model (2HDM) there are two electroweak doublets, in contrast to the one doublet of the SM \cite{Gunion:1989we}. The 2HDM can be of Types I, II or III depending on how the doublets couple to fermions. In Type I one doublet does not couple to fermions, in Type II one doublet ($H_{u}$) couples to up-type fermions  while the other doublet ($H_{d}$) couples to down-type fermions, and in Type III both doublets couple to both up- and down-type fermions. The Higgs sector of the 2HDM Type II contains five bosons: $A, h, H, H^{\pm}$ where $A$ is a neutral pseudoscalar, $h,H$ are neutral scalars and $H^{+},H^{-}$ are charged scalars. At tree level, only three parameters of the Higgs sector are free: $\cos \alpha$, which diagonalizes the neutral CP-even mass matrix, $m_{A}$, one of the Higgs masses, and $\tan \beta = \langle H_u \rangle / \langle H_d \rangle$, the ratio of the vacuum expectation values of the doublets.

Supersymmetry (SUSY) is one theoretical  model which can cancel the quadratic divergence of the Higgs mass by positing fermionic partners for every SM boson and bosonic partners for every SM fermion \cite{Martin:1997ns}. It also has a natural candidate for dark matter, the lightest SUSY partner (LSP). The Minimal SUSY Model (MSSM) contains only these partners, and employs a 2HDM Type II \cite{Gunion:1989we}. In the MSSM the $A$ coupling to down-type fermions $f_d f_d$ is proportional to $m_{f_{d}} \tan \beta$, while the coupling to up-type fermions $f_u f_u$ is proportional to $m_{f_{u}} \cot \beta$ \cite{Carena:2002es}. There is no tree level $A$ coupling to gauge bosons, so $A \rightarrow b\bar{b}$ is the dominant decay and $A \rightarrow \tau^+ \tau^-$ is subdominant, though the latter has a cleaner experimental signature at the LHC. Similarly the $H^{\pm}$ coupling to $f_d f_u$ is proportional to $m_{f_d} \tan \beta $ and $ m_{f_u} \cot \beta$ \cite{Carena:2002es}, so for  $\tan \beta >1$,  $H^+ \rightarrow \tau^+ \nu$ dominates for $m_{H^\pm} \lesssim m_{top}$,  while $H^+ \rightarrow t\bar{b}$ dominates for $m_{H^\pm} \gtrsim m_{top}$. LEPII has excluded $m_{A}<93$~GeV at 95\% C.L. \cite{Schael:2006cr}, and this constrains $m_{H^{\pm}}$ to lie above this scale as indicated by the tree level mass relation $m_{H^{\pm}}^2=m_{A}^2+m_{W}^2$ \cite{Carena:2002es}.

The Next-to-Minimal Supersymmetric Model (NMSSM) is motivated by the $\mu$-term problem of the MSSM \cite{Ellwanger:2009dp}. It solves the problem with an additional electroweak singlet $S$, and the resulting Higgs sector contains the MSSM Higgs content plus an additional neutral  pseudoscalar and an additional neutral scalar: $a_1, a_2, h_1, h_2, h_3, H^{\pm}$ where the subscripts denote mass ordering \cite{Ellwanger:2011sk}. In addition to the free parameter $\tan \beta$ of the MSSM, there are an additional five free parameters in the NMSSM Higgs sector: an effective $\mu$-term $\mu_{eff}=\sqrt{2} \langle S \rangle$, $\lambda$ and $A_{\lambda}$, which describe the coupling strength of the $S$ to the doublets $H_u$ and $H_d$, and finally $\kappa$ and $A_{\kappa}$, which describe the trilinear self-coupling of $S$ \cite{Ellwanger:2011sk}. In contrast to the $A$ of the MSSM, the $a_1$ of the NMSSM can be very light and therefore allows a Higgs sector phenomenology distinct from the MSSM \cite{Dermisek:2006wr}

In the following we review the current public 2HDM, MSSM, NMSSM and invisible Higgs decay searches at ATLAS. A full description of the ATLAS detector can be found in \cite{1748-0221-3-08-S08003}. All searches described use the CLs technique for setting limits \cite{0954-3899-28-10-313}.

\section{2HDM Higgs Searches}

The search for the heavy CP-even neutral $H$ of the 2HDM was carried out with 14fb$^{-1}$ at $\sqrt{s}=8$~TeV. Full details can be found in \cite{ATLAS-CONF-2013-027}. Signal $h/H \rightarrow WW^{\star}$ in $gg$ and VBF production modes and the dominant background, $WW^{\star}$, are simulated  with Powheg \cite{Nason:2010ap} and Pythia \cite{Sjostrand:2006za}.

A trigger requiring a single $e$ or $\mu$ with $p_{T}>24$~GeV is required. Exactly one $e$ and one $\mu$ with opposite charge are required, with the leading lepton $p_{T}>25$~GeV and the sub-leading lepton $p_{T}>15$~GeV. Relative track and calorimeter isolation is required for both $e$ and $\mu$ and the invariant mass $m_{e\mu}>10$~GeV. Finally $E_{T}^{miss}>25$~GeV for $\Delta \phi(j\ell, E_{T}^{miss})>\pi/2$ or $E_{T}^{miss} \times \sin \Delta \phi(j \ell,E_{T}^{miss})>25$~GeV otherwise ($j \ell$ indicates the closest to $E_{T}^{miss}$ in $\phi$ of $e, \mu$, or jet).

Events are separated into two categories, 0-jet and 2-jet. For 0-jet events, $\Delta \phi (e,\mu)<2.4$ and $m_{e \mu}<75$~GeV are required. For 2-jet events, $\eta_{j1} \times \eta_{j2}<0$, $m_{e \mu}<75$~GeV and the transverse mass $m_{T}^{e \mu}<180$~ are required with zero $b$-tags. After full selection, $1490 \pm 420$ events are expected in the 0-jet category, with 1815 observed, while $450 \pm 180$ events are expected in the 2-jet category, with 483 observed. Approximately 470 (76) signal ($m_{A}=150, \tan \beta=3, \alpha=\pi/2$) type I 2HDM signal events are expected  in the 0-jet (2-jet) channel.

Further discrimination between signal and background is provided by a multivariate neural network discriminant with kinematic inputs tailored to the 0-jet or 2-jet category. Exclusion limits in the $\cos \alpha$-$m_{H}$ plane are obtained after including systematic uncertainties with the CLs technique. For $\tan \beta=20$, ATLAS has excluded most points in the $\cos \alpha - m_{H}$ plane below $m_{H} \lesssim 200$~GeV for Type I and many points with $m_{H} \lesssim 200$~GeV for Type II.

Another 2HDM search documented in \cite{ATLAS-CONF-2013-081} selects for the channel $t \rightarrow c H$, a flavor changing neutral current process with SM branching ratio of $B_{SM}(t \rightarrow c H)=3 \times 10^{-15}$. In a type III 2HDM, for favorable regions of parameter space, the branching ratio can be as high as $B_{2HDM3}(t \rightarrow c H)=1.5 \times 10^{-3}$. With the 4.7fb$^{-1}$ integrated luminosity at $\sqrt{s}=7$~TeV and 20.3fb$^{-1}$ integrated luminosity at $\sqrt{s}=8$~TeV, the 95\% C.L. limit is established at $8 \times 10^{-3}$.

\section{MSSM Higgs Searches}

\begin{figure}[t]
\centering
\includegraphics[width=0.495\textwidth]{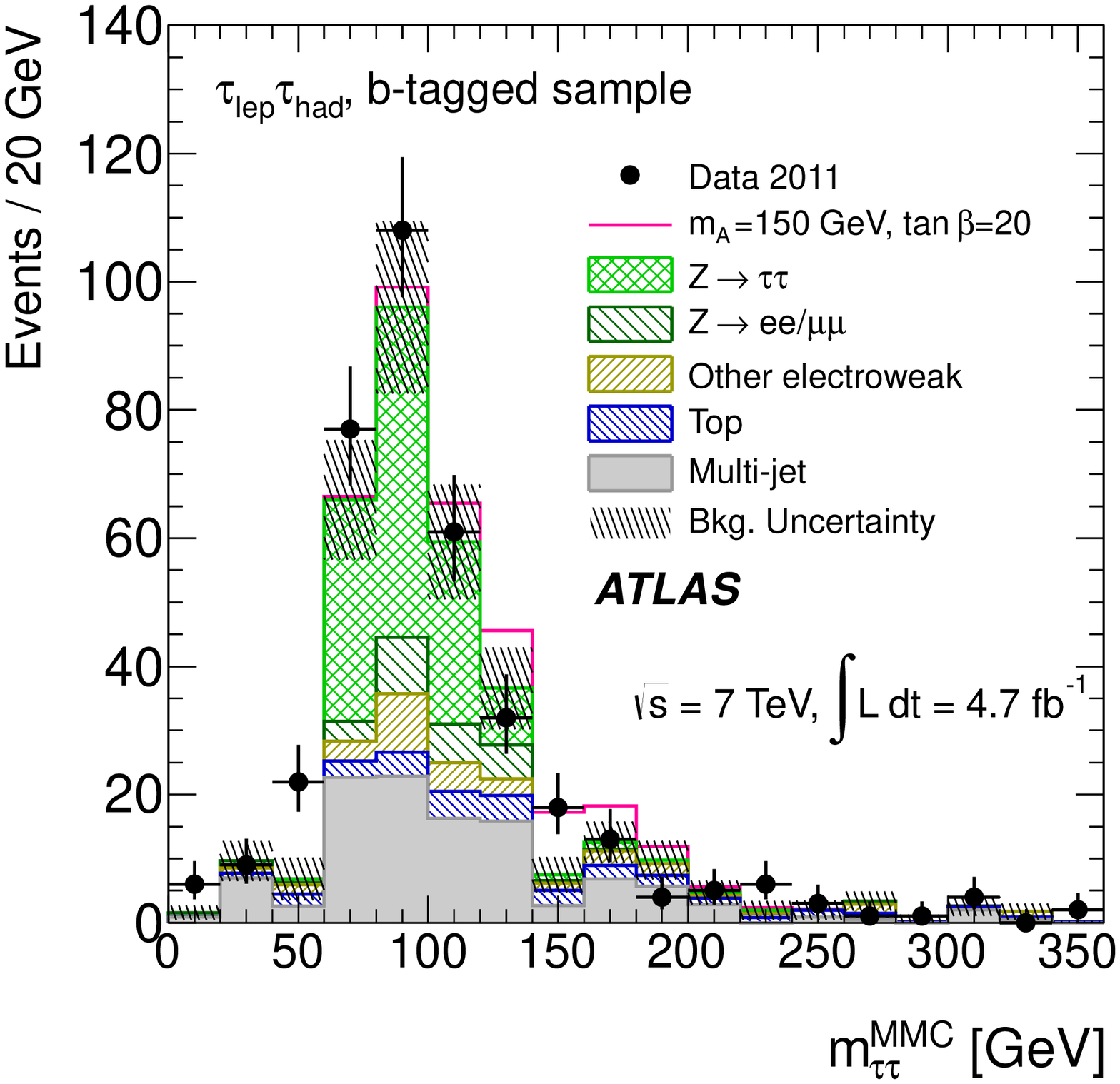}
\includegraphics[width=0.495\textwidth]{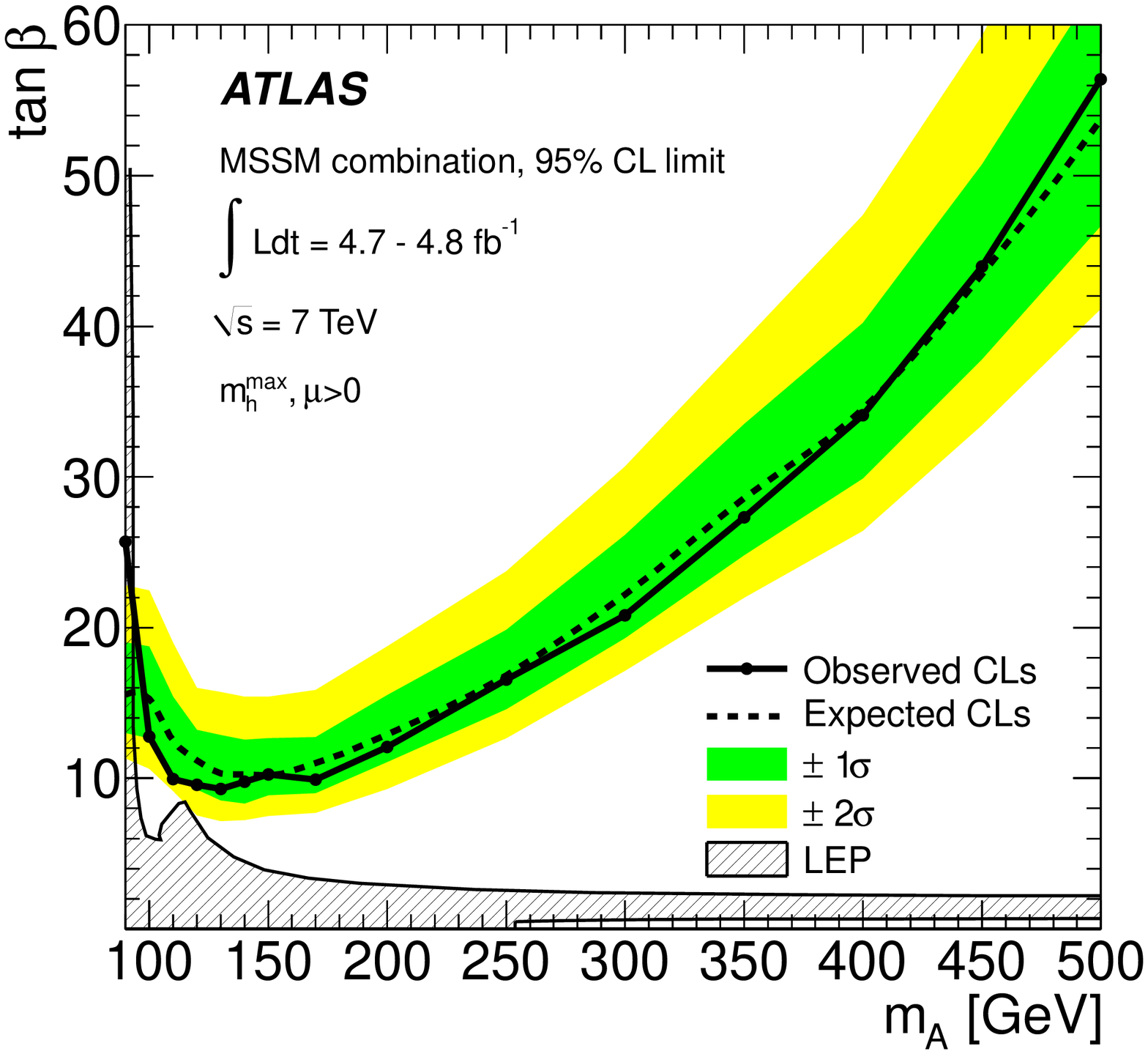}
\caption{For the MSSM neutral Higgs search, the reconstructed $\tau$ pair mass in the $\tau_{lep}\tau_{had}$ channel $b$-tagged sample (left) and the combined $\tau_{lep}\tau_{had}$, $\tau_{lep}\tau_{lep}$, $\tau_{had}\tau_{had}$ exclusion region in the $\tan \beta - m_{A}$ plane (right). }
\label{fig:tautau}
\end{figure}

We focus on the channels $H^+ \rightarrow \tau^+ \nu$ and $A/h/H \rightarrow \tau^+ \tau^-$, which have the greatest sensitivity at the LHC. ATLAS has also published a search for $H^+ \rightarrow c \bar{s}$ \cite{Aad:2013hla}, which has sensitivity for low mass and small $\tan \beta$.

\subsection{$A/h/H \rightarrow \tau^+ \tau^-$}

A general search for all neutral MSSM Higgs bosons with 4.7fb$^{-1}$ at $\sqrt{s}=7$~TeV, with focus on the $A \rightarrow \tau^+ \tau^-$ is documented in \cite{Aad:2012cfr}. We focus on the $\tau_{lep} \tau_{had}$  decay channel with most sensitivity and refer the reader to \cite{Aad:2012cfr} for the channels $\tau_{lep} \tau_{lep}$ and $\tau_{had} \tau_{had}$. The signal simulation samples use Sherpa \cite{Gleisberg:2008ta} for $b\bar{b}A$ production and Powheg for $gg$ production, while background simulation uses Alpgen \cite{Mangano:2002ea} and Pythia for $W/Z$+jets and MC@NLO \cite{Frixione:2010wd} for $t\bar{t}$.

Event selection begins by requiring a single $e$ or $\mu$ trigger and offline selection $p_{T}>20$~GeV with $\vert \eta  \vert < 2.5$ for $\mu$ or $p_{T}>25$~GeV with $\vert \eta \vert <2.47$ for $e$, on the efficiency plateau for both triggers. Both $e$ and $\mu$ are required to satisfy relative track and calorimeter isolation. Additional $e$ or $\mu$ are vetoed. One hadronic $\tau_{had}$ of opposite charge to the $e$ or $\mu$ is required, reconstructed from one- or three-tracks using a multivariate BDT technique which employs shower shape, isolation, and collimation information.

The dominant backgrounds are $Z \rightarrow \tau^+ \tau^-$ and $t\bar{t}$. In order to reduce background with W bosons, the transverse W mass is required to satisfy $m_{T}^{W}<30$~GeV. The $Z \rightarrow \tau^+ \tau^-$ background is modeled from a high purity $Z \rightarrow \mu^+ \mu^-$ sample in which the $\mu$ are removed from the event and are replaced with fully simulated $\tau$ decays. $W$+jets and $t\bar{t}$ background shapes are taken from simulation but normalized with data control samples. The QCD multijets background shape and normalization are both taken from data control samples.

The $A/h/H \rightarrow \tau^+ \tau^-$ mass is reconstructed using the technique described in \cite{Elagin:2010aw}. Events are split into two samples, one with a $b$-tagged jet and another with no $b$-tagged jet but $E_{T}^{miss}>20$~GeV. After full $b$-tag sample selection, $180 \pm 20$ events are expected in the $e$ channel, with 202 observed, while $154 \pm 30$ events are expected in the $\mu$ channel, with 175 observed.  For signal $b\bar{b}A$ simulation with $m_{A}=150$~GeV and $\tan \beta=20$, approximately 20 (15) events are expected in the $\mu$ ($e$) channel with $b$-tag. See Figure \ref{fig:tautau} for the reconstructed $A$ mass distribution in the $b$-tagged sample (left) and the derived limits in the $\tan \beta$-$m_{A}$ plane after incorporating systematic uncertainties using the CLs technique.

\subsection{$H^+ \rightarrow \tau^+ \nu$}

\begin{figure}[t]
\centering
\includegraphics[width=0.495\textwidth]{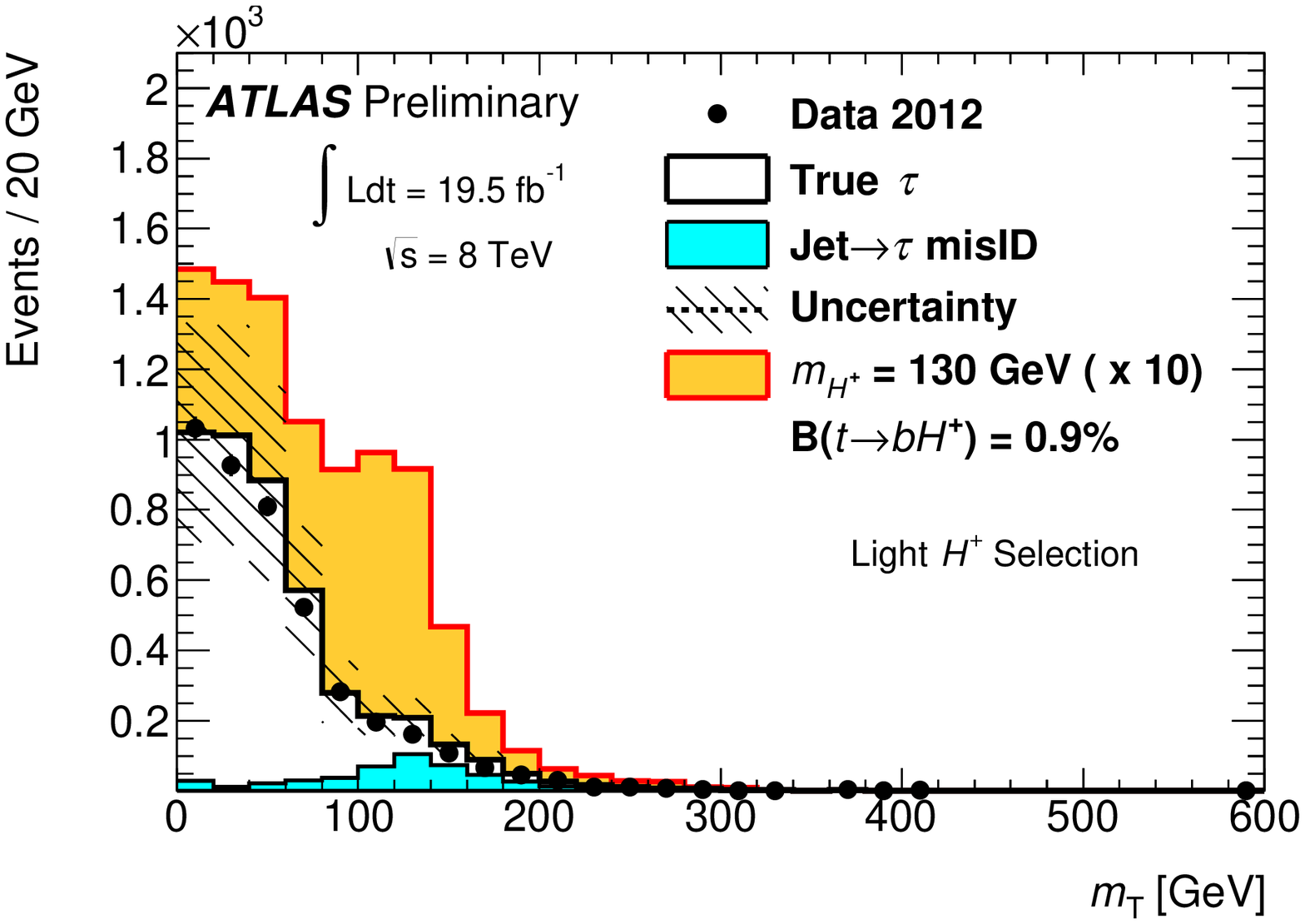}
\includegraphics[width=0.495\textwidth]{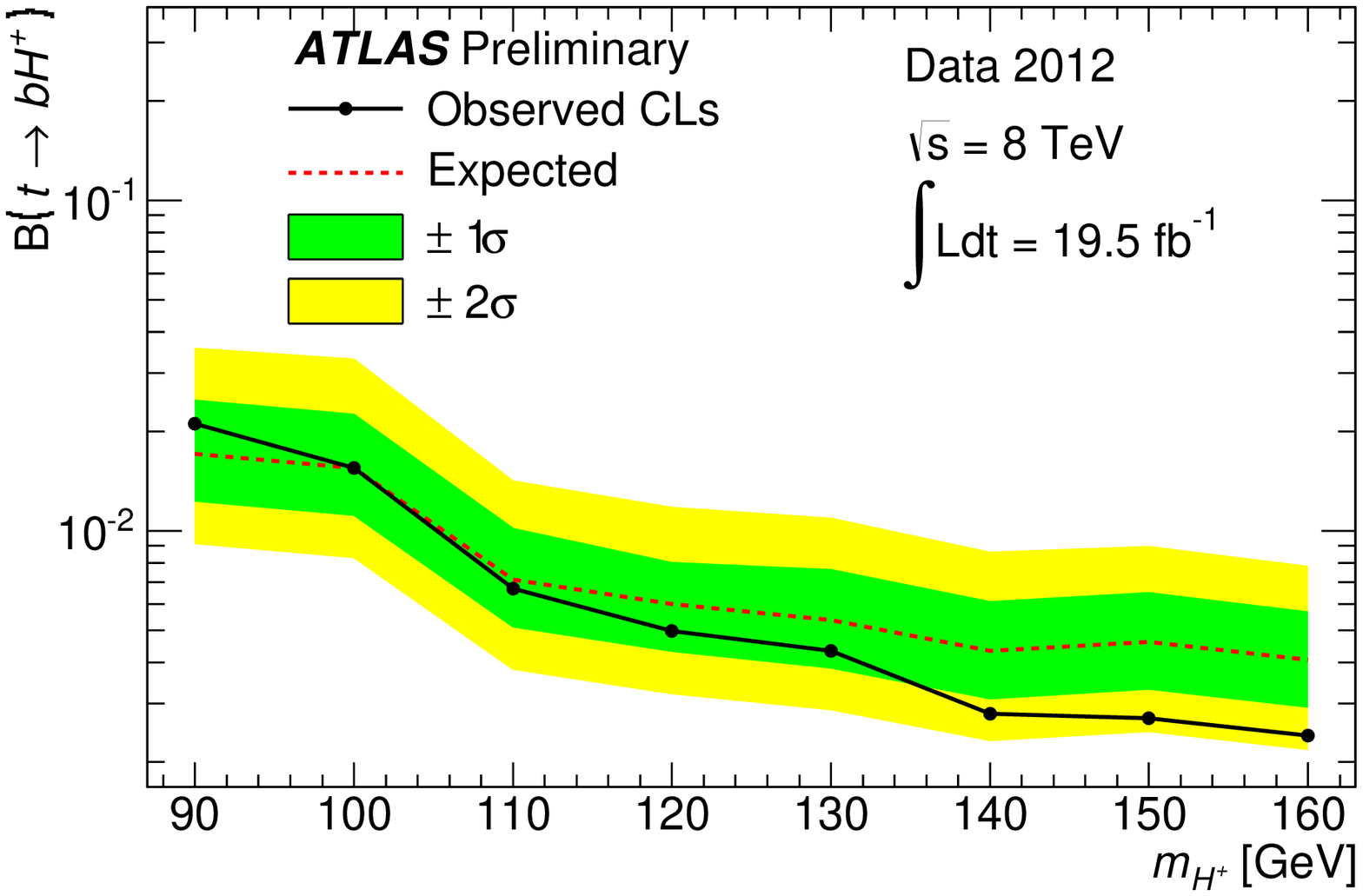}
\caption{For the MSSM $H^+ \rightarrow \tau \nu$ search, the $H^+$ transverse mass in data and signal simulation for $m_{H^+}=130$~GeV and $B(t \rightarrow bH^+)=0.9$\% (left) and the 95\% C.L. limit on $B(t \rightarrow bH^+)$.}
\label{fig:taunu}
\end{figure}

A search for $H^+ \rightarrow \tau^+ \nu$ with 19.5fb$^{-1}$ at $\sqrt{s}=8$~TeV is documented in \cite{ATLAS-CONF-2013-090}. Signal simulation is performed with Pythia for masses in the range $90 < m_{H^+} < 600$~GeV. We focus on the light $H^+$ in the $\tau_{had}$+jets channel, for which $m_{H^+}<m_{top}$ and sensitivity is highest.

For the light $H^+$, event selection is intended to identify top pair production, where one top decays via $t \rightarrow bH^+$ and the other via $t \rightarrow bW$. The subsequent decay $H^+ \rightarrow \tau^+ \nu$ is identified by hadronic $\tau_{had}$ reconstruction, while the decay $W \rightarrow qq^{\prime}$ is identified in two jets. Trigger selection requires either $E_{T}^{miss}>40$~GeV and a $\tau_{had}$ with $p_{T}>29$~GeV or $E_{T}^{miss}>50$~GeV and a $\tau_{had}$ with $p_{T}>27$~GeV. Four jets with $p_{T}>20$~GeV and $\vert \eta \vert <2.4$ are required, one of which must be $b$-tagged. One $\tau_{had}$ is required with $p_T>40$GeV and $E_{T}^{miss}>65$~GeV. Additional $\tau_{had}$ with $p_T>20$~GeV are vetoed, as are events with $e$ ($\mu$) with $p_T>25$~GeV ($p_T>20$~GeV). Finally $E_{T}^{miss}/\sqrt{\sum p_T}>26$~GeV$^{1/2}$ where the sum is over tracks in the primary vertex. See Figure \ref{fig:taunu} (left) for the reconstructed transverse $H^+$ mass $m_{T}(\tau_{had},E_{T}^{miss})$ after full selection.

Background from events with true $\tau$ is estimated from simulation while background from jets faking $\tau$ is estimated with a data-driven technique. The expected number of background events is $4500\pm 100 \pm 800$, while 4230 events are observed in data. For signal with $m_{H^+}=130$~GeV, approximately 500 events are expected. No excess is observed, and 95\% C.L. limits are placed in the $\tan \beta - m_{H^+}$ plane and on $B(t \rightarrow bH^+)$ assuming $B( H^+ \rightarrow \tau \nu)=1$. See Figure \ref{fig:taunu} (right) for limits on $B(t \rightarrow bH^+)$ vs $m_{H^{\pm}}$. This result excludes nearly all points in the $\tan \beta - m_{H^{\pm}}$ plane for $m_{H^{\pm}}<m_{top}$.

\section{NMSSM Higgs Searches}

\begin{figure}[t]
\centering
\includegraphics[width=0.495\textwidth]{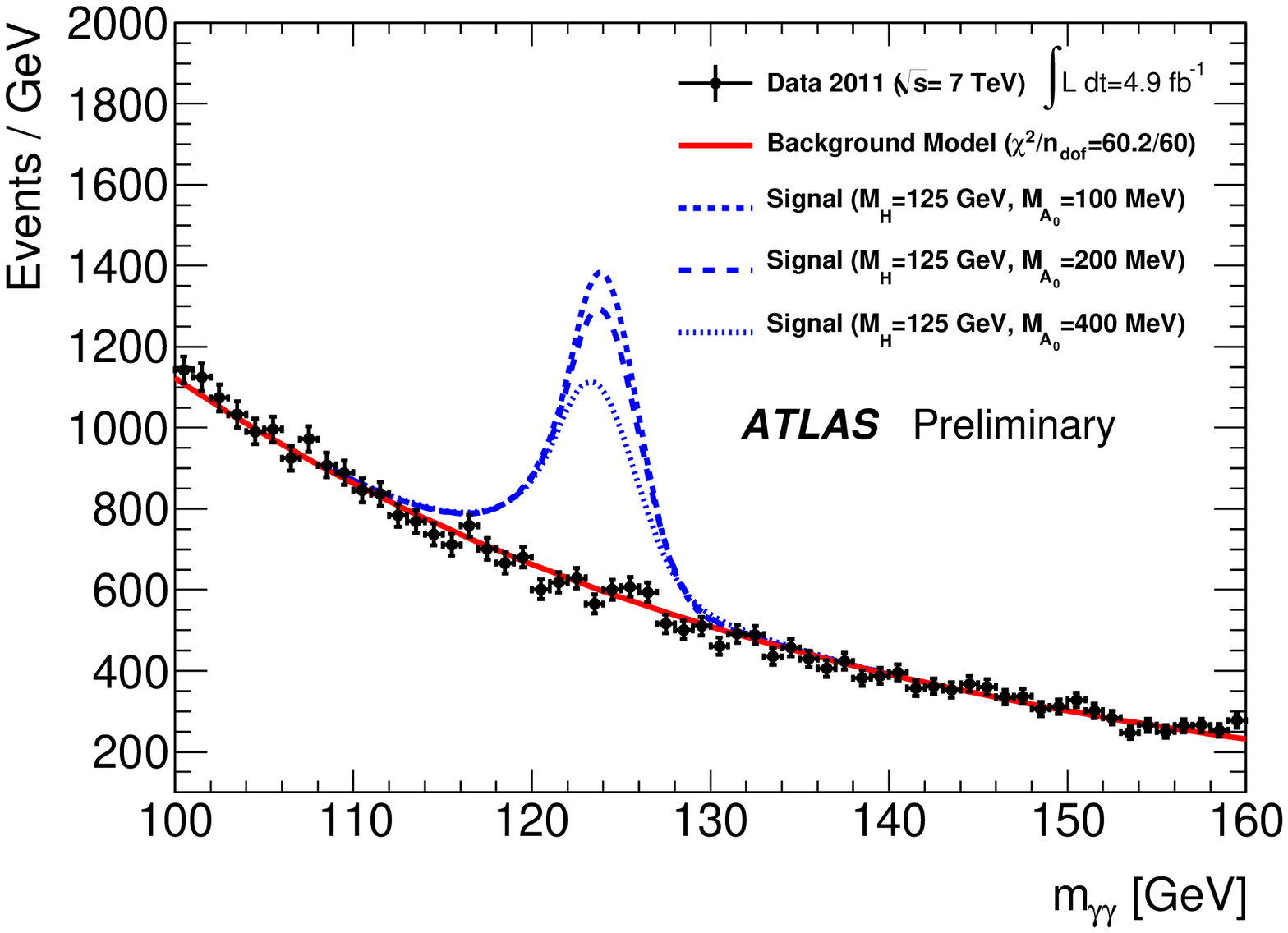}
\includegraphics[width=0.495\textwidth]{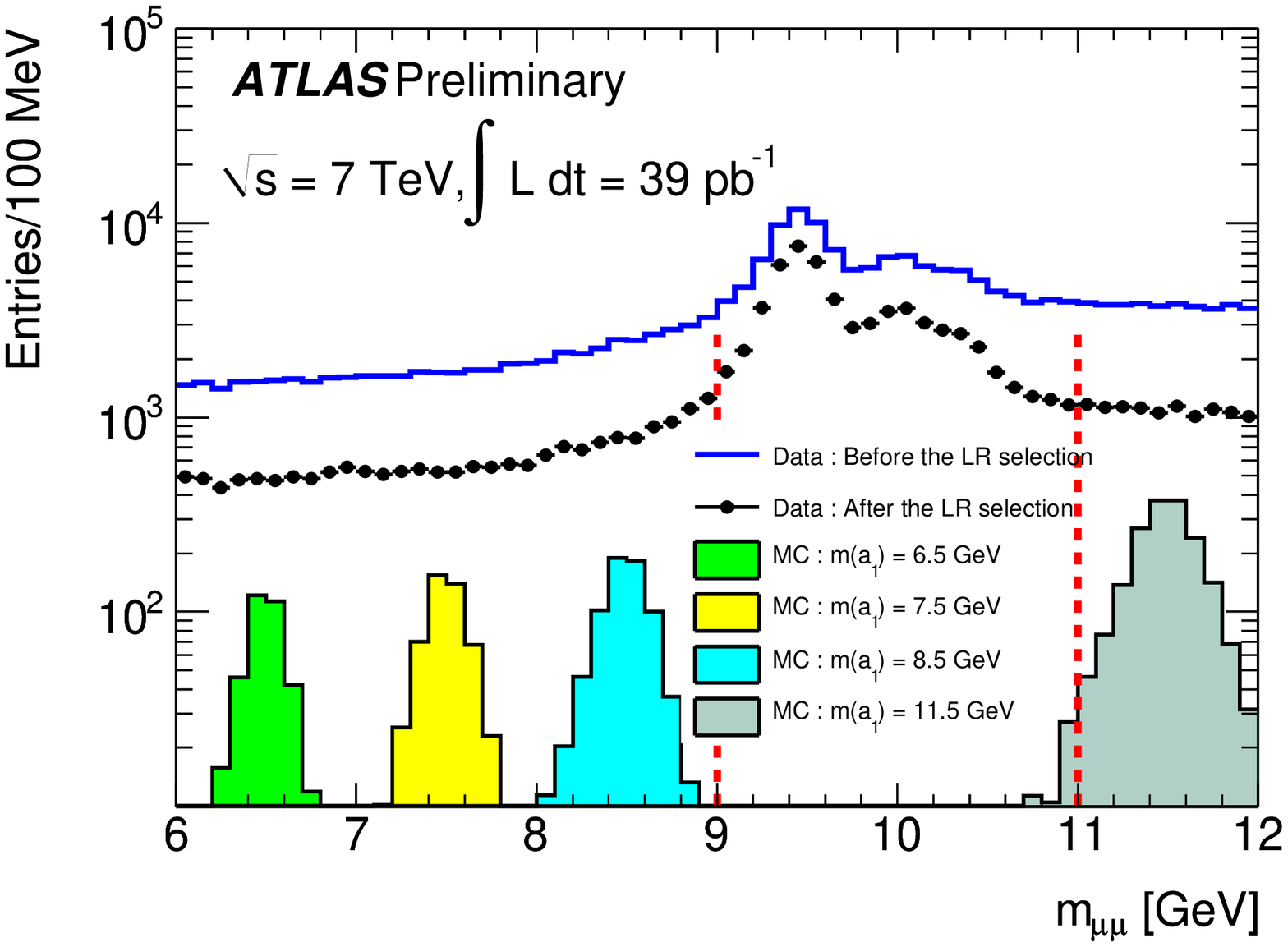}
\caption{Reconstructed mass for NMSSM $h_1 \rightarrow 2a_1 \rightarrow 4 \gamma$ (left) and  $a_{1} \rightarrow \mu^+ \mu^-$  (right). In signal $a_1 \rightarrow 2 \gamma$, each $\gamma$ pair are so highly collimated that they are reconstructed as a single $\gamma$.}
\label{fig:nmssm}
\end{figure}

\subsection{$h_{1} \rightarrow 2a_1 \rightarrow 4 \gamma$}

A search for $h_1 \rightarrow 2a_1 \rightarrow 4 \gamma$ with 4.9fb$^{-1}$ at $\sqrt{s}=7$~TeV is documented in \cite{ATLAS-CONF-2012-079}. Signal simulation is performed with Pythia for $m_{a_1}=100,200,400$~MeV and $110 < m_{h_1} < 150$~GeV while the background model is obtained from a data fit. For such light $a_1$, the $\gamma$ in $a_1 \rightarrow 2 \gamma$ are so highly collimated that they are typically reconstructed as a single $\gamma$.

Event selection begins with trigger selection, which requires two loose trigger $\gamma$ with $E_{T}>20$~GeV. Offline selection requires two tight $\gamma$ with $E_{T}>25$~GeV and calorimeter isolation. The signal trigger efficiency is estimated to be better than 96\%, while the offline selection efficiencies vary from 16\% to 30\% depending on $m_{a_1}$ and $m_{h_1}$. The background is modeled with a fit to the data distribution. See Figure \ref{fig:nmssm} (left) for the photon pair mass distribution in signal simulation and data. Observed 95\% C.L. limits of approximately 0.1pb are placed on the cross section for $pp \rightarrow h_1 \rightarrow 2a_1 \rightarrow 4\gamma$, with some variation depending on $m_{h_1}$ and $m_{a_1}$.

\subsection{$a_1 \rightarrow \mu^+ \mu^-$}

The search for direct production of a light $a_1$ decaying to $\mu$ pairs with 39.3pb$^{-1}$ at $\sqrt{s}=7$~TeV is documented in \cite{ATLAS-CONF-2011-020}. Signal simulation is performed with MC@NLO while background is estimated from data.

The event selection requires two oppositely charged offline $\mu$ with $p_{T}>4$~GeV and $\vert \eta \vert<2.5$ which match to $\mu$ which triggered the event with the same thresholds. The invariant mass of the $\mu$ pair is required to satisfy $4.5 < m_{\mu^+ \mu^-} < 14.0$~GeV. A likelihood is then constructed using the $\chi^2$ of the dimuon fit and the relative $\mu$ calorimeter isolation energy in order to discriminate from Drell-Yann and $\Upsilon$ background.  
See Figure \ref{fig:nmssm} (right) for the reconstructed $a_1 \rightarrow \mu^+ \mu^-$ mass in data before and after imposing a requirement on the likelihood distribution, together with simulated signal distributions. Observed 95\% C.L. limits of approximately 100pb are placed on the cross section for $gg \rightarrow a_1 \rightarrow \mu^+ \mu^-$, with some variation depending on $m_{a_1}$.

\section{Invisible Higgs Search}

In the SM, the only Higgs decay to an invisible final state is $H \rightarrow ZZ^{\star} \rightarrow \nu \bar{\nu} \nu \bar{\nu}$. This proceeds with a tiny branching ratio, but in BSM models invisible decay branching ratios can be large. For example, $h \rightarrow \chi^0 \chi^0$, where $\chi^0$ is the LSP in SUSY, is invisible if the $\chi^0$ is stable. Another example is an exotic decay like NMSSM $h_1 \rightarrow 2a_1 \rightarrow 4 \tau$, where the highly collimated $\tau$ escapes identification with standard reconstruction and the event satisfies the operational definition of \emph{invisible}.

The search for invisible Higgs decays with 4.7fb$^{-1}$ at $\sqrt{s}=7$~TeV and 13.0fb$^{-1}$ at $\sqrt{s}=8$~TeV is documented in \cite{ATLAS-CONF-2013-081}. Signal $q \bar{q} \rightarrow ZH \rightarrow \ell^+ \ell^- 4\nu$ ($\ell=e,\mu$) simulation is generated with Powheg and Herwig++ for mass in the range $115 < m_{H} < 300$~GeV. The dominant background is $ZZ \rightarrow \ell^+ \ell^- \nu \bar{\nu}$ and is simulated with Sherpa. 

Event selection requires a single or double $e$ or $\mu$ trigger. Events are required to have a pair of oppositely charged $e$ or $\mu$ with $p_{T}>20$~GeV and $76 < m_{\ell^+ \ell^-} < 106$~GeV, and are vetoed if there are any additional $e$ or $\mu$ with $p_{T}>7$~GeV. Furthermore $E_{T}^{miss}>90$~GeV, $\Delta \phi (E_{T}^{miss},p_{T}^{miss})<0.2$, $\Delta \phi (E_{T}^{miss},\ell^+ \ell^-)>2.6$ and $\vert E_{T}^{miss} -p_{T}^{\ell^+ \ell^-} \vert / p_{T}^{\ell^+ \ell^-} <0.2$ are required. Finally, the event is vetoed is there are one or more jets with $p_{T}>20$~GeV and $\vert \eta \vert < 2.5$. In the $\sqrt{s}=7$~TeV ($\sqrt{s}=8$~TeV) data, $32.7 \pm 1.0 \pm 2.6$ ($78 \pm 2.0 \pm 6.5$) events are expected while 27 (71) are observed. The 95\% C.L. expected (observed) upper limit on the branching ratio to invisible for $m_{H}=125$~GeV is 84\% (65\%), and the 95\% C.L. limits on cross section times branching ratio are found to vary in the approximate range $10 < \sigma \times BR < 30$~fb depending on $m_{H}$.

\section{Conclusion}

We have described the current state of the searches for BSM Higgs bosons at the ATLAS experiment. In a generic 2HDM Types I and II with $\tan \beta=20$, ATLAS has excluded most points in the $\cos \alpha - m_{H}$ plane below $m_{H} \lesssim 200$~GeV for Type I and many $m_{H} \lesssim 200$~GeV for Type II. The 95\% C.L. upper limit on $B(t \rightarrow cH)$, which can be enhanced by over enhanced in a 2HDM Type III over the SM by twelve orders of magnitude, is established at $8 \times 10^{-3}$. The ATLAS search for neutral MSSM decays to $\tau^+ \tau^-$ has excluded $\tan \beta \gtrsim 10$ for $m_{A} \lesssim 200$~GeV. The search for MSSM $H^+ \rightarrow \tau^+ \nu$ almost completely excludes $m_{H^{\pm}}<m_{top}$, with only small regions of $\tan \beta$ viable, and the 95\% C.L. upper limit on $B(t \rightarrow tH^+)$ ranges from 2\% to 0.2\%, depending on $m_{H^{\pm}}$, assuming $B(H^+ \rightarrow \tau^+ \nu)=1$. While the NMSSM searches have begun to constrain $\sigma \times BR$, the parameter space is much larger than in the MSSM, and constraining it has only just begun. Finally, the invisible Higgs search demonstrates that the 125 GeV boson may have a high decay rate to undetected invisible or exotic final states.

\section*{Acknowledgments}

Thanks to all colleagues past and present who have contributed to the construction and operation of the ATLAS experiment. Thanks also to colleagues at the LHC, whose effort made the searches described here possible.

\bibliography{paper}{}

\end{document}